\documentclass[aps,showpacs,preprintnumbers,amsmath,amssymb�eqsecnum,
twocolumn, tightenlines,
]{revtex4}

\usepackage[dvips]{graphicx}
\usepackage{bm}
\usepackage{longtable}
\usepackage{amsmath}
\usepackage{dcolumn}

\begin{document}

\bibliographystyle{apsrev}

\title {Extension of the Schiff  theorem to ions and molecules}

\author{V. V. Flambaum} \email[Email:]{flambaum@phys.unsw.edu.au}
\author{A. Kozlov} \email[Email:]{o.kozloff@student.unsw.edu.au}

\affiliation{School of Physics, University of New South Wales, Sydney
  2052, Australia}

    \date{\today}

    \begin{abstract}
According to the Schiff theorem the nuclear electric  dipole moment (EDM) is screened in neutral atoms. In ions this screening is incomplete. We extend  a derivation of the  Schiff theorem to  ions and molecules. The finite nuclear size effects are considered including $Z^2 \alpha^2$ corrections to the nuclear Schiff moment which are significant in all atoms and molecules of experimental interest. We show that in majority of  ionized atoms the  nuclear EDM contribution to the atomic EDM dominates  while  in  molecules the contribution of the Schiff moment dominates. We also consider the screening of  electron EDM in ions.
    \end{abstract}


    \maketitle

  \section{Introduction}
\indent
Permanent electric dipole moment (EDM) of elementary particle or  atom  violates
 both $P$ and $T$ invariance.
  The Kobayashi-Maskawa mechanism leads to extremely small values
of the EDMs of the particles. It is also too weak to explain the
matter-antimatter asymmetry of the Universe. On the other hand, most
of the popular extensions  predict much larger EDMs which
are within experimental reach. Therefore, measurements of EDM  provide
an excellent method to search for physics beyond the Standard Model. The measurements
of EDM in atomic and molecular experiments are presented in Refs. \cite{1,2,3,4,5,6,7,8,9,10,11,12,13,14,15,16,17,Leanhardt:2011,Workshop}.
 
  The EDM of an atom  is mostly due to
either electron EDM and T,P-odd electron-nucleon interactions in
paramagnetic systems (with non-zero electron angular  momentum $J$) or due to the
$T,P$-odd nuclear forces in diamagnetic systems
($J=0$; nuclear-spin-dependent e-N interaction contributes here too).
The existence of $T,P$-odd nuclear forces leads to the $T,P$-odd
nuclear moments in the expansion of
the nuclear potential in  powers of distance $R$ from the center of
the nucleus. The lowest-order term in the expansion, the nuclear EDM,
is unobservable in neutral atoms due to the total screening of the
external electric field by atomic electrons \cite{Schiff:1963}. It might be possible
however to observe the nuclear EDM in ions, where it is screened incompletely (see e.g. \cite{ Dzuba:1986,Itoi:2010,Dzuba:2010}). The first
non-vanishing terms which survive the screening in neutral systems are
the  Schiff moment which was defined  in Ref. \cite{Sushkov:1984} (see also Refs. \cite{Sandars:1967,Hinds:1980} where the contribution of the proton EDM was considered) and the electric octupole moment (the latter vanishes in nuclei of experimental interest  which have  spin 1/2). More accurate treatment of the finite nuclear size in Ref.
\cite{Flambaum:2002}
has shown that the atomic EDM is actually produced by the  nuclear Local dipole moment which differs  from the Schiff moment  by a correction  $\sim Z^2 \alpha^2$  where $Z$ is the nuclear charge and $\alpha$ is the fine structure constant. Since all experiments deal with heavy atoms this correction is significant.
 
In the non-relativistic classical limit the screening  formulas can be obtained in a very simple way. The
second Newton law for the ion and its nucleus in the electric field reads
\begin{align}\label{acceleration}
  (M_N+N_em_e) a_i & = (Z-N_e) e E_0\\
  M_N a_N & = Z e E_N\\
  m_e a_e & = e E_e,
\end{align}
where $m_e$ and $M_N$ are the electron and nuclear masses;   $a_i$, $a_N$ and $a_e$ are the ion, nucleus and  electron average accelerations
respectively, $E_0$ is the external electric field,  $E_N$ is the average electric field at the
nucleus, $E_e$ is the average electric field at one of the ion electrons, $e$ is the proton  charge, $N_e$ is the number of electrons in the ion. Since
system of particles moves altogether, the averaged accelerations must be equal ($a_i=a_N=a_e$),
therefore
\begin{align}
 E_N & = \frac{Z-N_e}{Z}E_0\frac{M_N}{M_N+N_em_e} \approx  (1-N_e/Z)E_0\\
 E_e & \approx  (Z-N_e)\frac{m_e}{M_N}E_0.\label{E_e}
\end{align}
As we can see,  the average electric field for electrons is suppressed by the ratio of masses $m_e/M_N$ that is very small for heavy atoms. It means that in the non-relativistic  limit there is  practically no effect related to the electron EDM in heavy atoms and ions, $-{\bf d_e \cdot E}_e \approx 0$. The interaction of the nuclear EDM $d$ with the external field, $-{\bf d \cdot E_{N}}$, is suppressed by the factor $(Z-N_e)/Z$.  \\
\indent The same approach can be used to determine the  electric field at the nucleus in a  diatomic molecule:
\begin{align*}
  (M_1+M_2+N_em_e) a_i & = (Z_1+Z_2-N_e) e E_0, \nonumber \\
  M_2 a_2 & = Z_2 e E_{2N}, \nonumber
\end{align*}
\begin{equation}\label{6}
 E_{2N}=\frac{Z_1+Z_2-N_e}{Z_2}\frac{M_2}{M_1+M_2+N_em_e} E_0.
\end{equation}
Screening is stronger for diatomic molecules because of  the factor $ M_2/(M_1+M_2)$ that contains both nuclear masses. This
	indicates that the nuclear motion can not be ignored. We also see that in neutral atoms and molecules the field at the nucleus is zero, therefore  the interaction of the nuclear  EDM $d$ with the screened electric field vanishes, ${\bf d E}_N$=0. Similarly, 
\begin{equation}\label{7}
 E_{e}=(Z_1+Z_2 - N_e)\frac{m_e}{M_1+M_2+N_em_e} E_0.
\end{equation}

The different  screening laws of EDM in neutral atoms, ions and molecules raise a number of new questions. For example, is the screening term in the  nuclear Schiff moment different in neutral atoms and ions? Can nuclear  motion in molecules produce any additional effects which do not exist in a single atom? Are there any new effects of the electron density polarization in ions and molecules?
Simple classical formulas presented above do not answer these questions. This motivates us to revisit the quantum Schiff theorem \cite{Schiff:1963} and extend it to  the cases of ions and molecules.   We also derive a formula which more accurately takes into account  the finite nuclear size and calculate corrections to the nuclear Schiff moment.

The present work is also motivated by new experiments.
Effects of EDM in molecules are enhanced \cite{Hinds:1980, Sandars:1967,Cho:1991,Sushkov:1978}.   This is why the molecular experiments are so popular now. Recently the EDM experiment
has been started with molecular ions  \cite{Leanhardt:2011}. The EDM experiments with atomic
ions in the storage rings  have been considered too \cite{Workshop}. 

  \section{Screening of EDM in atomic ions}
    \subsection{Nuclear EDM and Schiff moment}

\indent The charge distribution in a finite size nucleus can be written as $\rho({\bf r}) = \rho_0({\bf r}) + \delta\rho({\bf r})$, where $\int\rho_0d^3r=1$, $\delta\rho({\bf r})$ is due to  the $P,T$-odd  interactions. The $P,T$-odd term in charge density leads to the nonzero nuclear dipole moment ${\bf d} = d{\bf I}/I = Ze\int d^3r\delta\rho\,{\bf r}$, where $Ze$ is the nucleus charge, $e$ is the proton charge. Let us define $N_e$ as the number of electrons. If $N_e \ne Z$ a system is an ion. In a neutral atom  ($N_e=Z$) our derivation is expected to give the same results as the Schiff theorem \cite{Schiff:1963} including the effects of the finite nuclear size \cite{Sushkov:1984,Flambaum:2002}.\\
\indent The Hamiltonian of a single atom in an external electric field $E_0$ can be written in the following form: 
\begin{equation}\label{Ham1}
\hat H=\hat T + \hat V_0 + \hat V + \hat U + \hat W\text{ ,}
\end{equation}
where 
\begin{align*}
& \hat T=\sum_i^{N_e} \frac{-\hbar^2}{2m_e}\frac{\partial^2}{\partial{\bf R}_i\,^2}-\frac{\hbar^2}{2M_N}\frac{\partial^2}{\partial{\bf q}_N\,^2} \,,\\
& \hat V_0=\sum_{i>j}^{N_e}\frac{e^2}{|{\bf R}_i - {\bf R}_j|} - Ze^2\sum_i^{N_e}\int d^3r\frac{\rho_0(r)}{|{\bf R}_i - {\bf q}_N - {\bf r}|} \,,\\
& \hat V=\sum_i^{N_e}e{\bf R}_i{\bf E}_0 - Ze\,{\bf q}_N {\bf E}_0 ,\\
& \hat U=- Ze^2\sum_i^{N_e}\int d^3r\frac{\delta\rho({\bf r})}{|{\bf R}_i - {\bf q}_N - {\bf r}|}\,,\\
& \hat W=- {\bf d} {\bf E}_0 \,.
\end{align*}
Here ${\bf R}_i$ and ${\bf q}_N$ are the radius-vectors of the electrons and nucleus correspondingly. The expression for $\hat U$ can be expanded in powers of $r/R_i$ since the nuclei size is small compared to the atomic  scales. Let us keep the  first two nonvanishing terms:
\begin{align*}
\hat U = & -{\bf d} e \sum_i^{N_e} \frac{{\bf R}_i-{\bf q}_N}{|{\bf R}_i-{\bf q}_N|^3}\\
 & - 4\pi \frac{Ze^2}{10}\int d^3r\delta\rho r^2 {\bf r} \sum_i^{N_e}\nabla_i\delta({\bf R}_i-{\bf q}_N)\,.\\
\end{align*}
In the above expansion the octupole term was omitted since it leads to the mixing of the states with high electron angular momentum and its contribution to the total atomic  EDM is small \cite{Sushkov:1984}. 

Following Schiff let us define the operator
\begin{equation}
\hat Q=\frac{{\bf d}}{Ze}\frac{\partial}{\partial{\bf q}_N} \,.
\end{equation}
\\
It is easy to check  that there is a relation between $[\hat Q, \hat V_0]$ and $\hat U$
\begin{align}
& \hat U = \left[ \hat Q, \hat V_0\right] - 4\pi e{\bf S} \sum_i^{N_e}\nabla_i\delta({\bf R}_i-{\bf q}_N)\\
& {\bf S} = \frac{1}{10}\left\{ Ze\int d^3\delta\rho r^2 {\bf r} - \frac{5}{3}\,{\bf d}\int d^3r\rho_0(r)r^2\right\} \,,\label{SchiffMom}
\end{align}
where the expression for the Schiff moment ${\bf S}$ has the same form as for a neutral atom \cite{Sushkov:1984}. Substituting expression for $\hat U$ and $\hat W=\left[ \hat Q, \hat V\right]$ into Eq. (\ref{Ham1}) we obtain
\begin{equation}\label{Ham1Sym}
\hat H = \hat H_0 + \left[\hat Q, \hat H_0\right] - 4\pi e{\bf S} \sum_i^{N_e}\nabla_i\delta({\bf R}_i-{\bf q}_N) \,,
\end{equation}
where $\hat H_0 = \hat T + \hat V_0 + \hat V$ is the Hamiltonian of the system in the external electric  field without $P,T$-odd terms.
The calculation gives the following result for the commutator 
\begin{equation}\label{Acceler1}
\left[\hat Q , \hat H_0\right] = -\frac{\bf d}{Ze}\frac{i}{\hbar}\left[\hat H_0 , {\bf P}_N \right] = -\frac{\bf d}{Ze}M_N\hat{\bf a}_N \,,
\end{equation}
where $\hat{\bf a}_N$ is the nuclear acceleration operator. 
To obtain the average value of the acceleration operator we can use the Ehrenfest theorem:
\begin{equation}
\langle \hat {\bf a}_N\rangle = \frac{\langle {\bf F} \rangle}{M_N} = \frac{(Z-N_e)e\bf E_0}{M_N} \,,
\end{equation}
where $F$ is the average force acting on the nucleus (see Eq. (\ref{acceleration})).
Substituting the above expression to Eq. (\ref{Acceler1}) we obtain for the averaged commutator of $\hat Q$ and $\hat H_0$ the following equation
\begin{equation}
\langle \left[\hat Q , \hat H_0\right]  \rangle = -\left(1-\frac{N_e}{Z}\right){\bf d}{\bf E}_0 \,.
\end{equation}
Substituting this result  into Eq. (\ref{Ham1Sym}) we obtain  the effective Hamiltonian of the ion in the external electric field $E_0$:
\begin{equation}\label{Ham1Fin}
\hat H = \hat H_0 -\left(1-\frac{N_e}{Z}\right){\bf d}{\bf E}_0 - 4\pi e{\bf S} \sum_i^{N_e}\nabla_i\delta({\bf R}_i-{\bf q}_N) \,.
\end{equation}
 Note that the derivation above is done in the adiabatic approximation assuming that we can average over electron motion when we calculate the nuclear motion, i.e. we assume $m_e \ll M_N$.
 If the number of electrons $N_e = Z$ the EDM term in the above expression vanishes,
as the Schiff theorem predicts. In the ion case the nuclear EDM interacts with the average  field $E_N=E_0 (1-N_e/Z)$ that acts on the ion's nucleus. 

The last term in Eq. (\ref{Ham1Fin}),
\begin{equation}\label{Szero}
\hat H_w= -4\pi e{\bf S} \sum_i^{N_e}\nabla_i\delta({\bf R}_i-{\bf q}_N),
\end{equation}
induces the ion EDM directed along the nuclear spin (which is the direction of the nuclear Schiff moment ${\bf S}$),  similar to the EDM of neutral  atoms. This expression is not applicable for heavy atoms where the Dirac equation gives infinite results for the electron wave functions at the point-like nucleus. 
 Accurate account of the finite nuclear size  gives the following form for the corrected Schiff moment electrostatic potential (defined by $\hat H_w= -e \varphi_S({\bf R})$):
\begin{equation}\label{Finite}
\varphi_S({\bf R})=-\frac{3{\bf S'}\cdot {\bf R}}{B}\rho_0(R),
\end{equation}
where $B=\int \rho_0(R)R^4dR$ is the normalization constant. In the limit of the  point-like nucleus the expression (\ref {Finite}) agrees with Eq. (\ref{Szero}). The corrected Schiff moment  ${\bf S'}$ is given by the equation (see Appendix)
\begin{equation}
\begin{split}
&{\bf S'}=\frac{Ze}{10}\frac{1}{1-\frac{5}{14}Z^2\alpha^2}\cdot\\
&\left\{\left[\langle{\bf r}r^{2}\rangle-\frac{5}{
3}\langle{\bf r}\rangle\langle r^{2}\rangle-\frac{2}{3}\langle r_i\rangle\langle q_{ij}\rangle\right]\right.\\
&\left.-\frac{5}{28}\frac{Z^2\alpha^2}{R_N^2}\left[\langle{\bf r}r^{4}\rangle-\frac{7}{3
}\langle{\bf r}\rangle\langle r^{4}\rangle-\frac{4}{3}\langle r_i\rangle\langle q_{ij}r^2\rangle\right]\right\}
\end{split}
\end{equation}
where $q_{ij}$ is the quadrupole moment tensor.  Here we omitted higher order terms which are proportional to a small factor   $Z^4\alpha^4/9$.
 Outside the nuclear radius  $R_N$ the nuclear density $\rho_0(R)=0$ and  the potential (\ref{Finite}) vanishes in agreement with the Schiff theorem. Near the origin $\rho_0(R)=const$
and  the potential (\ref{Finite})  is a linear function of ${\bf R}$. Therefore,  the gradient of the Schiff moment potential (\ref {Finite}) gives a constant electric field inside
the nucleus which is directed along the nuclear spin. This electric field polarizes the electron distribution and produces the atomic EDM. 
The calculations of the atomic EDM have been  performed, for example,  in Refs.  \cite{Sushkov:1984,Dzuba:2000,Porsev:2011,Dzuba:2002}.

       Below we make rough estimates to compare the nuclear EDM  and the Schiff moment
contributions to the atomic EDM. In the case of a spherical nucleus the nuclear EDM $d$, the nuclear  Schiff moment  and the  atomic EDM $D_A$ induced by the Schiff moment have  been estimated in Ref.  \cite{Sushkov:1984}: 
\begin{equation}
d \sim  10^{-21}\eta e \cdot cm \,,
\end{equation}
\begin{equation}
D_A \sim  (Z/100)^2 \cdot 10^{-24} \eta e \cdot cm \,,
\end{equation}
where $\eta$ is the  strength constant of the nuclear $P,T$-odd interaction (in units of the weak
Fermi constant $G$). 
 Assuming the single
ionization we get for the nuclear EDM screening factor $1-N_e/Z=1/Z$. As a result,
for the ionic EDM induced by the nuclear EDM we get
the estimate $1/Z\cdot 10^{-21} \eta |e|$cm. Thus, for the spherical nuclei  the nuclear EDM contribution exceeds
the nuclear Schiff moment contribution by at least one order of magnitude. However, in  heavy ions containing nuclei with the octupole deformation (e.g. $^{225}$Ra$^+$ and  $^{223}$Rn$^+$) the Schiff moment contribution is enhanced by three  orders of magnitude  \cite{Auerbach:1996,
Auerbach:1997} and is comparable to the nuclear EDM contribution (which is also slightly enhanced in these ions).

    \subsection{Electron EDM}
\indent  For neutral atoms the electron EDM problem was investigated in \cite{Sandars:1965} and further developed in \cite{Flambaum:1976}. The Hamiltonian of the nucleus and relativistic electrons in the external electric field $E_0$ can be presented as
\begin{align}
\hat H & = \hat H_0 + \hat H_w \,,\\
\hat H_0 & = -\hbar^2\triangle_N/2M_N-Ze{\bf q}_N{\bf E}_0 + \nonumber\\
& \sum_{i}^{N_e}-i\hbar c\,{\bf \alpha}_i\nabla_i+ \beta_i mc^2-\frac{Ze^2}{|{\bf R}_i - {\bf q}_N|}+\nonumber\\
& \; e{\bf R}_i{\bf E}_0 + \sum_{j>i}\frac{e^2}{|{\bf R}_i - {\bf R}_j|}\\
\hat H_w & = -d_e\sum_{i}^{N_e}\beta_i{\bf\Sigma}_i{\bf E}_t \,,\\
{\bf \Sigma} & = \begin{pmatrix}
{\boldsymbol \sigma} & 0\\
0 & {\boldsymbol \sigma}\\
\end{pmatrix}\nonumber
\end{align}
where ${\bf E}_t$ is the total electric field acting on the electron which includes the external field
${\bf E}_0$, the nuclear field and the field of other electrons, ${\bf \alpha}$ and $\beta$ are  the Dirac matrices. It is convenient to present  $H_w$ as the sum of two terms
\begin{align}
\hat H_w & = \hat H_{1d}+\hat H_{2d}\,,\\
\hat H_{1d} &=-d_e\sum_{i}^{N_e} {\bf\Sigma}_i{\bf E}_t \,,\\
\hat H_{2d} &=-d_e\sum_{i}^{N_e}(\beta_i-1) {\bf\Sigma}_i{\bf E}_t \,.
\end{align}
As it was pointed in  \cite{Sandars:1965} the  first term $H_{1d}$ gives no contribution to atomic EDM in a neutral atom. In an ion the $H_{1d}$ contribution is suppressed by a small factor $m_e/M_N$. It can be  shown using the commutator relation 
\begin{align}
\hat H_{1d} & = \left[\hat Q, \hat H_0\right] \,,\\
\hat Q & = -\frac{d_e}{e}\sum_i^{N_e}{\bf \Sigma}_i\frac{\partial}{\partial{\bf R}_i}.
\end{align}
Note that the matrix elements of the operators  in the the $H_{1d}$  come from the atomic size area where valence electrons 
(which contribute to the atomic angular momentum and EDM) are non-relativistic.
To estimate the average value of the commutator $\left[\hat Q, \hat H_0\right]$ the Erehnfest theorem can be employed
\begin{align}
\langle\left[\hat Q, \hat H_0\right]\rangle & =\frac{d_e}{e}\langle\sum_i{\bf \Sigma}_i\frac{{d\bf p}_i}{d t}\rangle\nonumber\\
\langle\sum_i{\bf \Sigma}_i\frac{{d \bf p}_i}{d t}\rangle & \approx \langle\sum_i{\bf \Sigma}_i{\bf F}_i\rangle =-e\langle\sum_i{\bf \Sigma}_i{\bf E}_e\rangle
\end{align}
Substituting expression (\ref{E_e}) for ${\bf E}_e$ into above equation we obtain for the average value of  $\hat H_{1d}$
\begin{equation}\label{H1d}
\langle\hat H_{1d}\rangle \approx -d_e\frac{m_e}{M_N}(Z-N_e)\langle\sum_i{\bf \Sigma}_i{\bf E}_0\rangle
\end{equation}
We see that the averaged value $\langle\hat H_{1d}\rangle$ is suppressed by the small mass ratio $m_e/M_N$. It means, that in the limit of heavy nucleus $\hat H_{1d}$ gives no contribution to EDM.  \\
\indent The second perturbation term $\hat H_{2d}$ vanishes in the non-relativistic limit since the
matrix $(\beta_i-1)$ acts on the lower components of the Dirac 4-spinors only. The operator $\hat H_{2d}$  induces atomic EDM  given by the same expression as for neutral atoms, except for  the sum in the matrix elements is taken over electron number $N_e<Z$:
\begin{equation}\label{elecEDM}
\begin{split}
{\bf D}_2 = &\, d_e\langle 0|\sum(\beta_i-1){\bf \Sigma}_i|0\rangle+\\
& 2e d_e\sum_n\frac{\langle 0|\sum(\beta_i-1){\bf \Sigma}_i{\bf E}_t|n\rangle\langle n|\sum{\bf R}_i|0\rangle}{E_0-E_n}
\end{split}
\end{equation}
In heavy atoms  the major contribution to $D_2$ comes from the second term ($D_2 \sim 3 R_{rel} Z^3 \alpha^2 d_e$ where $R_{rel} \sim 3$ is the relativistic factor  \cite{Sandars:1965,Flambaum:1976}) . This term
corresponds to the atomic  EDM due to the perturbation of the electron density by the operator 
  $\hat H_{2d}$.  Note that a similar equation with the perturbation  $\hat H_{1d}$ gives zero result due to exact cancellation between the first and second terms.  Indeed,
the zero and the first order corrections to the atomic EDM induced by $\hat H_{1d}$ give EDM  
\begin{equation}\label{elecEDM1}
\begin{split}
{\bf D}_1 = &\, d_e\langle 0|\sum{\bf \Sigma}_i|0\rangle+\\
& e\sum_n\frac{\langle 0|\left[\hat Q, \hat H_0\right]|n\rangle\langle n|\sum {\bf R}_i|0\rangle}{E_0-E_n}+\\
&  e\sum_n\frac{\langle 0|\sum {\bf R}_i|n\rangle\langle n|\left[\hat Q, \hat H_0\right]|0\rangle}{E_0-E_n}
\end{split}
\end{equation}
The above expression can be simplified in the following way. For the matrix elements of the commutators the following relations are valid
\begin{align} 
\langle n|\left[\hat Q, \hat H_0\right]|0\rangle=-(E_0-E_n)\langle n|\hat Q|0\rangle\\
\langle 0|\left[\hat Q, \hat H_0\right]|n\rangle=(E_0-E_n)\langle 0|\hat Q|n\rangle
\end{align}
Substituting these expressions into Eq. (\ref{elecEDM}) and using the completeness condition $\sum|n\rangle\langle n|=\hat 1$ we obtain
\begin{equation}
\begin{split}
& {\bf D}_1 =e\sum_n\sum_i\left[\langle 0|\hat Q|n\rangle\langle n|{\bf R}_i|0\rangle-\langle 0|{\bf R}_i|n\rangle\langle n|\hat Q|0\rangle\right]+\\
& d_e\langle 0|\sum{\bf \Sigma}_i|0\rangle=d_e\langle 0|{\bf \Sigma}_i|0\rangle+\sum_i e\langle 0|\left[\hat Q, {\bf R}_i\right]|0\rangle
\end{split}
\end{equation}
Using  definition of the operator $\hat Q$ it is easy to show that $\left[\hat Q, {\bf R}_i\right]=-d_e/e{\bf\Sigma}_i$. Hence, the second term in the above equation cancels the first term, so the dipole moment ${\bf D}_1$ induced by $H_{1d}$ equals to zero.   In this  derivation we assume that the electron states are stationary. This is valid if we neglect the ion acceleration. Therefore, the result  is consistent  with Eq. (\ref{H1d}).\\ 
\indent We see that  EDM of an ion induced by the electron EDM is given by the same equation (\ref{elecEDM}) as for neutral atoms (up to corrections $\sim m_e/M_N$).  A similar conclusion is also valid for molecular ions.

  \section{Nuclear EDM and Schiff moment in molecular ions}

\indent Let us consider a  molecular ion with $N_e$ electrons and two nuclei with charges $Z_1e$ and $Z_2e$. We assume that the second nucleus has EDM  ${\bf d}$ and Schiff moment ${\bf S}$. The molecular  Hamiltonian is equal to the sum of the following terms:
\begin{align*}
& \hat T = \sum_i^{N_e} \frac{-\hbar^2}{2m_e}\frac{\partial^2}{\partial{\bf R}_i\,^2} - \frac{\hbar^2}{2M_1}\frac{\partial^2}{\partial{\bf q}_1\,^2}-\frac{\hbar^2}{2M_2}\frac{\partial^2}{\partial{\bf q}_2\,^2} ,&\\
& \hat V_0 = \sum_{i>j}^{N_e}\frac{e^2}{|{\bf R}_i - {\bf R}_j|} - Z_2e^2\sum_i^{N_e}\int d^3r\frac{\rho(r)}{|{\bf R}_i - {\bf q}_2 - {\bf r}|} &\\
& - \sum_i^{N_e}\frac{Z_1e^2}{|{\bf R}_i - {\bf q}_1|}+Z_1Z_2e^2\int d^3r\frac{\rho(r)}{|{\bf q}_1 - {\bf q}_2 - {\bf r}|}, &\\
& \hat V = \sum_i^{N_e}e{\bf R}_i{\bf E}_0 - Z_1e\,{\bf q}_1 {\bf E}_0-Z_2e\,{\bf q}_2 {\bf E}_0, &\\
& \hat U = - Ze^2\sum_i^{N_e}\int d^3r\frac{\delta\rho({\bf r})}{|{\bf R}_i - {\bf q}_2 - {\bf r}|}, &\\
& + Z_1Z_2e^2\int d^3r\frac{\delta\rho({\bf r})}{|{\bf q}_1 - {\bf q}_2 - {\bf r}|}, &\\
& \hat W = - {\bf d} {\bf E}_0 ,&\\
\end{align*}
where ${\bf q}_1$ and ${\bf q}_2$ are the coordinates of first and second nuclei respectively. Using the operator  
\begin{equation}
\hat Q = \frac{{\bf d}}{Z_2e}\frac{\partial}{\partial{\bf q}_2}
\end{equation}
we can present the molecular  Hamiltonian in the form  similar to Eq. (\ref{Ham1Sym}):
\begin{align}\label{Ham2Sym}
\hat H = &\,\hat H_0 + \left[\hat Q, \hat H_0\right]\\
&- 4\pi e{\bf S} \left\{\sum_i^{N_e}\nabla_i\delta({\bf R}_i-{\bf q}_2)-Z_1\frac{\partial}{\partial{\bf q}_1}\delta({\bf q}_1-{\bf q}_2)\right\}.\nonumber
\end{align}
To calculate the average value of the commutator $\hat Q$ and $\hat H_0$ we can use the same algorithm as for a single atom. 
\begin{equation}\label{Acceler2}
\left[\hat Q , \hat H_0\right] =  -\frac{\bf d}{Z_2e}\frac{i}{\hbar}\left[\hat H_0 , {\bf P_2} \right] = -\frac{\bf d}{Z_2e}M_2\hat{\bf a}_2
\end{equation}
Since the molecule moves as a single body the average accelerations of all its particles is equal to the molecular acceleration, i.e.
\begin{align}
\langle \hat {\bf a_2}\rangle = \frac{\langle {\bf F} \rangle}{M_1+M_2+N_em_e} \approx \frac{(Z_1+Z_2-N_e)e\bf E_0}{M_1+M_2},\\
\langle \left[\hat Q , \hat H_0\right]\rangle  = -\frac{M_2}{M_1+M_2}\frac{Z_1+Z_2 -N_e}{Z_2}{\bf d}{\bf E}_0.
\end{align}
Finally, the effective Hamiltonian of the  molecular ion is
\begin{align}\label{Ham2Fin}
\hat H = \,& \hat H_0 -\frac{M_2}{M_1+M_2}\frac{Z_1+Z_2 -N_e}{Z_2}{\bf d}{\bf E}_0\\
& - 4\pi e{\bf S} \left\{\sum_i^{N_e}\nabla_i\delta({\bf R}_i-{\bf q}_2)-Z_1\frac{\partial}{\partial{\bf q}_1}\delta({\bf q}_1-{\bf q}_2)\right\},\nonumber
\end{align}
Thus,  in a molecular ion the EDM   term experiences the extra
suppression. As for the Schiff moment term, it is
still described by the same operator as for a single atom, except
for the extra term proportional to $\partial(\delta({\bf q}_1-{\bf
q}_2))/\partial{\bf q}_1$ describing the interaction of the charge of the first nucleus and the Schiff moment of the second nucleus. The matrix elements of such interaction are extremely small due to the  Coulomb barrier.

  \section{Enhancement of the Schiff moment contribution to $P,T$-odd effects in polar molecules}
\indent Now we can compare the contributions of the nuclear EDM and Schiff moment to $P,T$-odd effects in polar molecular ions. Important difference between  molecules  and single
atoms is that the nuclear motion significantly affects induced $P,T$-odd effects. The  Schiff moment  contribution  in polar molecules is  enhanced because of the strong internal electric field \cite{Sandars:1967}.
 Another interpretation of the enhancement  is due to the small distance between the opposite parity rotational  levels \cite{Sushkov:1978,Sushkov:1984}.   

The nuclear  $P,T$-odd effects are studied in the molecules with zero electron angular momentum. After averaging
Hamiltonian Eq. (\ref{Ham2Fin}) over electron wave function we obtain the effective Hamiltonian for the nuclear motion: 
\begin{equation}
\hat H = -\frac{\hbar^2}{2\mu}\triangle_q + U_e +
\frac{\mu\omega^2}{2}(q-q_e)^2 + BJ(J+1) + \hat H_w
\end{equation}
where ${\bf q}={\bf q}_1 - {\bf q}_2$, $q_e$ is the equilibrium
distance between the nuclei in averaged potential, J is the
rotational angular momentum of the molecule, $U_e$ describes the interaction of the partially screened nuclear EDM,  the Schiff moment term $\hat H_w$ can be presented as \cite{Hinds:1980,Sushkov:1984}
\begin{equation}\label{mol}
\hat H_w = 6  X S \frac{{\bf I \cdot n}}{I},
\end{equation}
where ${\bf S} = S{\bf I}/I$, ${\bf n}$ is  the unit vector along the molecular axis, $X$ is the constant that appears after averaging the
perturbation over the  electron wave function. In the first order of the perturbation theory the Schiff term  leads to the
rotation state mixing
\begin{equation}
\psi^{(1)}=6 X S \frac{I_z}{I}\sum_{J'\ne
J}\frac{\langle Jm|n_z|J'm\rangle}{E_J-E_{J'}}|J'm\rangle
\end{equation}
where $\psi^{(0)}=|Jm\rangle$ is the unperturbed rotational  wave function.
Since the energy difference $E_J-E_{J'}=B\{J(J+1) - J'(J'+1)\}$
can be very small for rotation levels, the state mixing can be
significant. This mixing induces EDM in the rotational state
\begin{align}
D_z^S = 2\langle\psi^{(0)}|D_M n_z|\psi^{(1)}\rangle \\
 =\frac{6  X S D_M
I_z}{IB}\frac{J(J+1)-3m^2}{J(J+1)(2J-1)(2J+3)}\nonumber\\
 \equiv K_m S I_z/I \,.
\end{align}
Here ${\bf D}_M = D_M{\bf n}$ is the internal EDM of the polar molecule.
This formula is valid for  $J\ne 0$. For $J=0$ the induced EDM is
\begin{equation}\label{MolK}
D_z^S = -\frac{2  X S D_M I_z}{IB} \equiv K_m S I_z/I
\end{equation}
\indent There is also the screened nuclear EDM  contribution $D_z^d$ to $P,T$-odd molecular EDM ( see  Eq. (\ref{Ham2Fin})). Combining this contribution with the Schiff moment contribution
$D_z^S$ we obtained the $P,T$-odd part of the interaction of a molecular ion with the external electric field $E_0$: 
\begin{equation}\label{V}
V = -\left(\frac{M_2}{M_1+M_2}\frac{Z_1+Z_2 -N_e}{Z_2} d - K_m S\right)\frac{{\bf I}{\bf E}_0}{I}
\end{equation}
This equation  tells us that there is actually no enhancement of the electric field in the polar  molecule since the electric field at the nucleus is suppressed $1/Z_2$ times rather than enhanced. However, there is huge enhancement of the Schiff moment contribution since the expression for the coefficient $K_m$ contains in the denominator the  rotational constant $B$ which may be  five orders of magnitude smaller than the interval between  atomic levels of opposite parity.

 Note that we can derive Eq. (\ref{V}) treating $E_0$ as a perturbation. Therefore,  the energy shift produced by the Schiff moment  in Eq. (\ref{V}) is actually proportional to the average polarization of the polar molecule in the electric field $E_0$. In the small electric field
it is linear in $E_0$, however, in the high field it tends to the constant. This determines the saturation effect in the energy shift produced by the Schiff moment  if we go  beyond the weak electric field $E_0$ approximation (see Eq. (\ref{mol}) where the average polarization $n_z<1$) . 

\indent Using  Eq. (\ref{V})  we can compare molecular  EDM induced by the screened nuclear  EDM  and the Schiff moment. Consider, for example, molecule PbF$^+$ since it has the same number of electrons as a well studied  molecule TlF  where the effect of the nuclear Schiff moment has been measured. The  screened EDM  term for PbF$^+$ is  $D_N\sim  10^{-23} \eta e \cdot $cm ( EDM of F and EDM of  odd isotope of Pb give comparable contributions  since values of $M/Z$ are approximately the same).  To obtain the Schiff moment induced EDM in the ground state we need to estimate the constant $K_m$, given by Eq. (\ref{MolK}). Since the molecular parameters are unknown for the  ion we assume them to be of the order of their values for the neutral molecule TlF:  $X\approx 8000$ a.u. \cite{Dzuba:2002, Petrov:2002}. The values of the  rotational  constant $B=1.025\cdot 10^{-6} $a.u. and  dipole moment $D_M=1.65$ a.u.  for TlF are  taken from \cite{Huber:1979}. Finally, substituting all the parameters into Eq. (\ref{MolK}) we obtain $K_m= 5 \cdot 10^{10}$ a.u. Assuming the Schiff moment value for an odd isotope of Pb equal to $S= 10^{-8}  \eta e \cdot fm^3$ \cite{Sushkov:1984} we obtain the  value for the   Schiff moment contribution $D_S  \sim 10^{-20}\eta e \cdot  \text{cm}$ which is three orders of magnitude larger than the nuclear  EDM contribution $D_N  \sim 10^{-23}\eta e \cdot  \text{cm}$.
As it was mentioned above, in  the nuclei with the octupole deformation like Ra$^{225}$ the Schiff moment is enhanced. Therefore,  in molecular ions like RaF$^+$ the Schiff moment induced EDM will be 5 orders of magnitude  larger than the partially  screened nuclear EDM. 

\section{Conclusions}
\indent Accurate treatment of the electron EDM effects shows that the T,P-odd EDM of atomic and molecular ions  at high $Z$  are dominated by the $Z^3$  enhanced relativistic  correction effect, similar to neutral systems. The direct contribution of electron EDM is suppressed
by the screening factor $(m_e/M)$ where $M$ is the  ion mass. 

 The situation is different for the nuclear EDM. In atoms the nuclear EDM is screened by the factor $Z_i/Z$ where  $Z_i$ is the ion charge.  However, the nuclear EDM still  dominates over the Schiff moment 
 induced atomic EDM (with exception of  heavy ions which contain nuclei with the octupole deformation like $^{225}$Ra   and $^{223}$Rn where the Schiff moment is strongly enhanced).
 
In  molecular ions the nuclear EDM screening is slightly stronger  than in atomic ions,
 the screening factor is $(M_N/M)(Z_i/Z)$. At the same the Schiff moment contribution is  enhanced $\sim M_N/m_e \sim  10^5$ times due to the mixing of the close rotational states of opposite parity.
There is the  additional  Schiff moment enhancement in such  molecular ions like RaF$^+$.   As a result, the Schiff moment contribution is $10^3-10^5$ times larger than  the screened nuclear EDM contribution.   

This combination of the large enhancement factors  makes molecular ion experiments an attractive alternative to the atomic EDM experiments.

\section{Appendix}
\indent According to Eq. (\ref{Ham1Fin}) in the limit of the point-like nucleus the Schiff moment potential and its matrix element are given by 
\begin{align}\label{grad}
\varphi_S({\bf R})&=4\pi{\bf S}\cdot{\boldsymbol\nabla}\delta({\bf R})\\
\langle s|-e\varphi_S|p\rangle&=4\pi e {\bf S}\cdot({\boldsymbol\nabla}\psi_s^{\dagger}\psi_p)_{R=0}
\end{align}
 For the solutions of the  Dirac equation $({\boldsymbol\nabla}\psi_s^{\dagger}\psi_p)_{R\rightarrow0}$ is infinite for a point-like nucleus. Therefore,  for relativistic electrons it is necessary to account for the finite size of the nucleus and  introduce a finite-size Schiff moment potential. An appropriate potential has been shown \cite {Flambaum:2002} to increase linearly inside the nucleus and vanish at the nuclear surface:  
 \begin{equation}\label{Finite1}
\varphi_S({\bf R})=-\frac{3{\bf S'}\cdot {\bf R}}{B}n(R),
\end{equation}
where $B=\int n(R)R^4dR\approx R_N^5/5$, $R_N$ is the nuclear radius and $n(R)$ is a smooth function which is 1 for $R<R_{N}-\delta$ and 0 for $R>R_{N}+\delta$; $n(R)$ can be taken as proportional to the nuclear density $\rho_0$ (note that we can choose any normalization of $n(r)$ since the normalization constant cancels out in the ratio $n/B$, see Eq. (\ref{Finite1})). 

Below we will accurately derive expression for the corrected Schiff moment $S'$ that corresponds to the potential (\ref{Finite1}). \\
\indent The $P,T$-odd part of the nuclear electrostatic potential with electron screening taken into account can be written in the following form (see e.g. \cite{Auerbach:1997} for the derivation):
\begin{equation}\label{ScrPot}
\varphi ({\bf R})=Z\int\frac{e\rho({\bf r})}{|{\bf R}-{\bf r}|}d^3r + {\bf d\cdot\nabla}\int\frac{e\rho({\bf r})}{|{\bf R}-{\bf r}|}d^3r
\end{equation}
 As it was shown in \cite{Flambaum:2002} the expansion of the Coulomb potential in (\ref{ScrPot}) in terms of the  Legendre polynomials gives the following dipole term in the  potential:
\begin{equation}\label{phi1}
\varphi^{(1)}({\bf R})=Ze{\bf R}\int_R^{\infty}\left(\frac{\langle{\bf r}\rangle}{R^3}-\frac{{\bf r}}{R^3}+\frac{{\bf r}}{r^3}+\frac{\langle r_i\rangle q_{ij}}{r^5}\right)\rho({\bf r})d^3r
\end{equation}
We see that $\varphi^{(1)}({\bf R})=0$ if $R>R_N$ (nuclear radius) since $\rho({\bf R})=0$ in that region. Therefore, corresponding matrix elements will depend on the electron wave functions behavior inside the nucleus.
\indent All the electron orbitals for $l>1$ are extremely small inside the nucleus. Therefore, we can limit our consideration to the matrix elements between $s$ and $p$ Dirac orbitals. We will use the following notations for the electron wavefunctions:
\begin{equation}
\psi({\bf R}) =\begin{pmatrix}
f(R)\Omega_{jlm}\\
-i({\boldsymbol \sigma}\cdot {\bf n})g(R)\Omega_{jlm}
\end{pmatrix}
\end{equation}
where $\Omega_{jlm}$ is a spherical spinor, ${\bf n=R}/R$, $f(R)$ and $g(R)$ are the radial functions. Using  $({\boldsymbol \sigma}\cdot{\bf n})^2=1$ we can write the electron transition density as
\begin{align}\label{Usp}
& \rho_{sp}({\bf R})=\psi_s^{\dagger}\psi_p=\Omega_{s}^{\dagger}\Omega_{p}U_{sp}(R)\\
U_{sp}(R)= & f_s(R)f_p(R)+g_s(R)g_p(R)=\sum_{k=1}^{\infty}b_kR^k
\end{align}
The expansion coefficients $b_k$ can be calculated analytically \cite{Flambaum:2002};  the summation is carried over  odd powers of $k$. Using Eqs. (\ref{phi1},\ref{Usp}) we can find the matrix elements of the electron-nucleus interaction,
\begin{multline}\label{almostlast}
\langle s|-e\varphi^{(1)}({\bf R})|p\rangle =-Ze^2\langle s|{\bf n}|p\rangle\cdot
\left\{\int_{0}^{\infty}\left[ \left(\langle {\bf r}\rangle-{\bf r}\right)\cdot\right.\right.\\
\left.\left.\int_{0}^{r}U_{sp}dR+\left(\frac{{\bf r}}{r^3}+\frac{\langle r_i\rangle q_{ij}}{r^5}\right)\int_{0}^{r}U_{sp}R^3dR\right]\rho d^3r\right\}=\\
-Ze^2\langle s|{\bf n}|p\rangle\cdot\left\{\sum_{k=1}^{\infty}\frac{b_k}{k+1}\left[\langle{\bf r}\rangle\langle r^{k+1}\rangle-\frac{3}{k+4}\langle{\bf r}r^{k+1}\rangle\right.\right.\\
\left.\left.+\frac{k+1}{k+4}\langle r_i\rangle\langle q_{ij}r^{k-1}\rangle\right]\right\},
\end{multline}
where $\langle s|{\bf n}|p\rangle=\int\Omega_s^{\dagger}{\bf n}\Omega_pd\phi\sin\theta d\theta$, $\langle r^n\rangle=\int\rho({\bf r})r^nd^3r$. Note, that all vector values $\langle {\bf r}r^n\rangle$ are due to $P,T$-odd correction $\delta \rho$ to the nuclear charge density $\rho_0$, while $\langle r^n\rangle$ are the usual $P,T$-even moments of the charge density starting from the mean-square radius $\langle r^2\rangle=r_q^2$ for $k=1$.\\
\indent We now set the matrix elements (\ref{almostlast}) of the true nuclear $T,P$-odd potential to be equal to the matrix elements of the equivalent potential   (\ref{Finite1}) which are given by 
\begin{equation}\label{last}
\begin{split}
\langle s|-e\phi({\bf R})|p\rangle&=15e\langle s|{\bf n}|p\rangle\cdot\frac{{\bf S'}}{R_N^5}\int_0^{\infty}U_{sp}R^3n(R)dR\\
&=15e\langle s|{\bf n}|p\rangle\cdot{\bf S'}\sum_{k=1}^{\infty}b_k\frac{R_N^{k-1}}{k+4},
\end{split}
\end{equation}
where we have made approximation $\int n(R)R^kdR\approx R_N^{k+1}/(k+4)$. Equating (\ref{almostlast}) and (\ref{last}) we obtain
\begin{equation}\label{Sprime}
\begin{split}
&{\bf S'}=\frac{Ze}{15}\frac{1}{\sum_{k=1}^{\infty}\frac{b_k}{b_1}\frac{1}{k+4}R_N^{k-1}}\sum_{k=1}^{\infty}\frac{b_k}{b_1}\frac{1}{k+1}\\
&\left[\frac{3}{k+4}\langle{\bf r}r^{k+1}\rangle-\langle{\bf r}\rangle\langle r^{k+1}\rangle-\frac{k+1}{k+4}\langle r_i\rangle\langle q_{ij}r^{k-1}\rangle\right]
\end{split}
\end{equation}
\indent Thus we have a possibility of separating the nuclear and electronic parts of the calculation of atomic EDMs. The nuclear calculation involves only the determination of ${\bf S'}$ and the atomic calculation involves only the effects produced by  the equivalent potential  (\ref{Finite1}).

 Note that $S'$ in eq.  (\ref{Sprime}) is different from the Local dipole moment $L$ defined in Ref. \cite{Flambaum:2002}: $L$ does not contain the sum in the denominator. The reason for the difference is that here we reduce the problem to the nuclear size effective potential   (\ref{Finite1})
 while in Ref. \cite{Flambaum:2002} the problem was reduced to the contact effective potential  (\ref{grad}) located in the center of the nucleus. 
 
  In the non-relativistic case $(Z\alpha\rightarrow 0)$ we have just $b_1\ne 0$, and
\begin{equation}
\lim_{Z\alpha\rightarrow0}{\bf S'}=\frac{Ze}{10}\left[\langle{\bf r}r^2\rangle-
\frac{5}{3}\langle{\bf r}\rangle\langle r^2\rangle-\frac{2}{3}\langle r_i\rangle\langle q_{ij}\rangle\right].
\end{equation}
This is the usual expression for the Schiff moment ${\bf S}$. In practice it may be sufficient to use only the first and third terms in the expansion of $U_{sp}$. In this case we need only  the ratio $b_3/b_1$. This ratio is different  for the  matrix elements $s$ - $p_{1/2}$ 
($b_3/b_1=-(3/5) Z^2\alpha^2/R_N^2$)  and  $s$ - $p_{3/2}$ ($b_3/b_1=-(9/20) Z^2\alpha^2/ R_N^2$).  However, with the 10\% accuracy we can use the  average of these two values $b_3/b_1\approx -0.5 Z^2\alpha^2/R_N^2$.

\end{document}